# Clinical Validation of myOSLchip system for Radiotherapy Dosimetry

H. Davis[1], J. Siebers[1], K. Wijesooriya[1], M. Mistro[1]

1. University of Virginia Department of Radiation Oncology, Charlottesville, VA

Email: yzx6tz@uvahealth.org (H. Davis)

**Abstract -** In radiation oncology, inter-fractional dosimetry is often performed with luminescent dosimeters to verify the accurate delivery of a plan and ensure patient safety. Optically stimulated luminescent detectors (OSLDs) are the most commonly used detector type which offers good dose linearity and accuracy in the megavoltage energy range. Freiberg Instruments offer a dosimetry system under the brand name myOLSchip which consists of a BeO OSL dosimeter, reader, and eraser. A Varian Truebeam was used to characterize the detectors and calibrate their response in order to perform in-situ dosimetry during treatment. The OSLDs were tested with both photon and electron beams from 6-15 MV and 6-20 MV respectively. The dose signal to dose conversion in this investigation follows the recommendations of TG-191 in developing a dose response curve and creating a batch calibration factor for each dosimeter. The repeatability of this system is also investigated following successive erasing and re-irradiation cycles. The results of this data have been compared to the stated accuracy and precision of the BeO detectors by the manufacturer and shown to have good dose linearity and repeatability across multiple exposures and erasure cycles.

**Introduction –** In modern radiotherapy, in-situ dosimetry is a valuable tool to ensure patient safety and plan quality. Many modern techniques require in-situ dosimetry in clinically relevant situations. This imposes a challenge for any dosimetric system to have a high sensitivity across a wide range of delivered dose and a convenient design as to not hinder the delivery of treatment or patient setup. Thermoluminescence dosimeters (TLD), diodes, film, and Metal-Oxide-Semiconductor Field Effect Transistors (MOSFETs) are commonly used for these types of measurements because of their combination of form-factor and good dosimetric qualities. TLD's have been widely used in radiation therapy but their difficulty in readout due to heating and annealing has impacted their continued use. Both diodes and MOSFETs are great detector designs for in-situ dosimetry due to their high precision and accuracy, however radiation damage is a concern causing signal degradation across their lifetime [1], [2], [3].

Optically stimulated luminescent detectors (OSLDs) have been widely used in the medical physics community for many years due to their ease of use, including ability to be placed in nearly arbitrary locations, signal stability, response linearity, and ability to be rapidly read out in the clinic. Like TLDs, OSLDs store charge when radiation interacts in the crystal lattice of the active material. This is because the valance and conduction bands of the crystal lattice are separated by a few eV. Impurities in the crystal introduced through dopants, or inherit to the lattice, act as luminescence centers where separated charges from the conduction and valance band can recombine. Depending on the "depth" of these recombination centers, or the relative energy difference between the center and the conduction/valence bands, this recombination of separated charge can take seconds to many years [4], [5]. During stimulation, charges are freed from traps by optical stimulation of the lattice by a visible spectrum light (instead of heat with TLDs) whereby the charge can recombine at luminescence centers and produce visible light shown in Figure 1. The stimulation occurs for a short amount of time and then a photomultiplier tube (PMT) is used to count the number of freed photons post stimulation. The amount of charge stored and therefore light released upon stimulation is proportional to the dose received by the device allowing for accurate dosimetry when calibrated under controlled conditions.

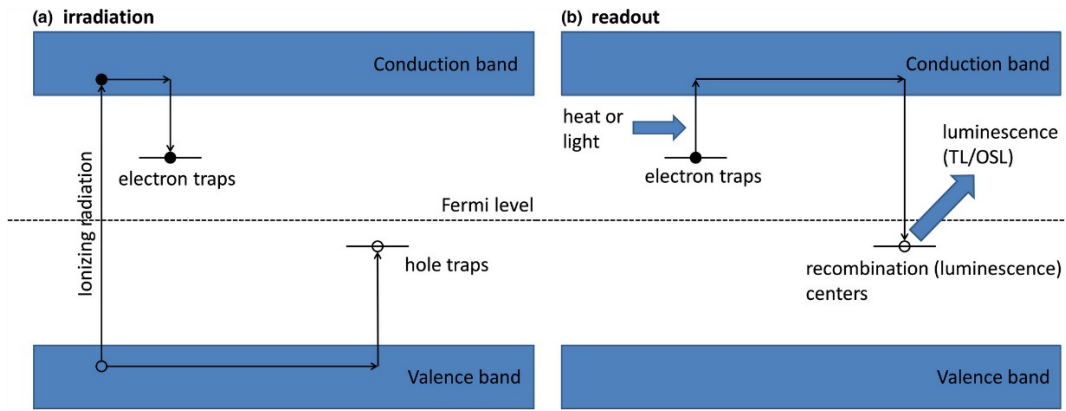

Figure 1. Energy level diagram for a OSL/TLD dosimeter showing (a) the separation of charges into traps for electrons and holes with (b) showing the stimulation of the trapped charges recombining at luminescence centers producing light [4].

The Landauer nanodot™ is one such OSLD comprised of $Al_2O_3$:C and was widely popular in clinical radiotherapy with many clinics using this device as an in-vivo dosimeter. The small form factor, light-tight encapsulation, and lowcost per dosimeter made the nanodot a very common device for a wide range of clinical dosimetry needs [6]. However, the nanodot is no longer available and has left users in dire need of a replacement system.

This study characterizes myOSL, an OSLD system that is now available in the US. Like the nanodot, myOSL has a small form factor, is packaged in a light-tight encapsulation, and has a low cost per dosimeter. AAPM TG191 references many publications which studied the nanodot system. However, the clinical performance of the RadPro myOSL system has hereto not been quantified [4]. This work investigates the dosimetric characteristics of myOSLchip, a beryllium oxide (BeO) based OSLD system. Linearity, accuracy, and other properties are explored in a clinical setting for therapeutic MV photon and electron beam dose measurements.

**Methods and Materials**

**Dosimeter -** The myOSLchip dosimeter (Figure 2.) is comprised of a 4.7 x 4.7 x 0.5 mm BeO element inside a 9.5 x 10 x 2 mm ABS (Acrylnitril-Butadien-Styrol-Copolymer) light-tight housing. The element has a density of 2.85 g/cm$^3$ with a $Z_{eff}$ of 7.21 [4]. Each chip has a 4 digit number and 2-D barcode printed on the housing for identification visually as well as by the reader. The BeO element is tightly fixed inside a tray that slides into the housing and is held in place with a pressure fit. The tolerances are sufficient to ensure the element will not inadvertently come out of the light tight housing outside of the reader or erasure devices. RadPro reports a coefficient of variation less than 1% and linearity better than 1% up to 10 Sv. A delay between exposure and readout of 10 minutes is recommended to allow shallow trap emission and signal stability. RadPro provides a sensitivity correction factor for each device upon request. For high dose rate applications, RadPro suggests a maximum single dose of 10,000 mGy, a lifetime dose of 5,000,000 mGy and a maximum accumulated bleaching time of 10,000 s.

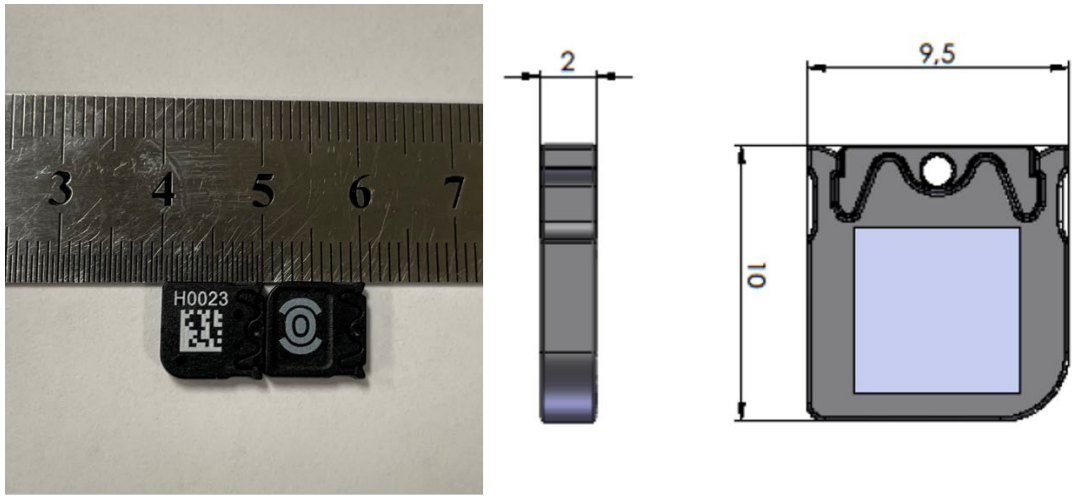

Figure 2. Left shows the printed 2-D bar code on the myOSLchip with a cm ruler for scale. Right shows a technical drawing with the dimensions in mm for scale with a grey/blue box showing the size of the BeO element inside.

**Reader** - The myOSLreader shown in Figure 3. is comprised of a Peltier element, photomultiplier tube, and readout electronics with an LCD screen on the device. There is a 100 mW read LED for stimulation and a 1 W high power erase LED all with 460 nm wavelength light which is centered in the spectral response curve of the BeO element. The Peltier element is to stabilize the chip's temperature at 25 C for consistent thermal response during stimulation. Readings are performed by placing the chip in the scalloped tray with the 2-D bar code facing outward and the tray-opening pin in the corresponding hole on the chip. After rotating the mechanism 90 degrees into the read position, the device opens the light-tight housing and provides the number counts detected after stimulation and store the value for the chip on the reader. The stimulation pulse length is 0.2 s with a 0.03 s delay followed by a 0.001 s read time to determine the number of photons released. Up to 100 dosimeters and 500 data readings can be stored directly on the device. A computer running the myOSLchip software is necessary for more devices. Rotating the mechanism an additional 45 degrees to the erase position will expose the chip to a high-power erase LED if clearing the signal is desired, but not a nessisary step in the readout process.

**Eraser -** The myOSLeraser is a standalone device that holds up to 48 myOSLchips in machined slots in a tray. A lever must be actuated when placing the tray into the housing of the eraser in order to open all of the chips, exposing the BeO elements to the high power LEDs. The user can select any erasure time in minutes up to 120 minutes which is sufficient to reduce the residual count from a 1050 cGy exposure of 240,000 counts to near background levels of less than 100 counts. When exposing many dosimeters to high doses exceeding the mGy – Gy range, the standalone eraser is necessary to fully bleach 48 dosimeters at once whereas the myOSLreader can only bleach one detector at a time.

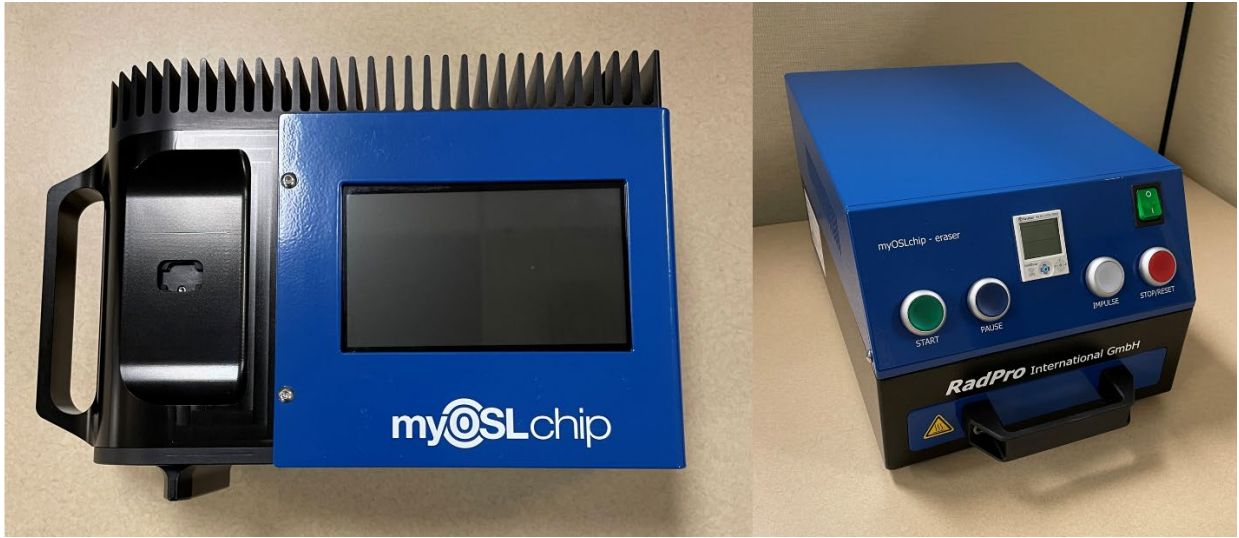

Figure 3. Left image shows the myOSLchip reader with the scalloped tray and slot for the dosimeter. The right image shows the high power LED erasure with a drawer capable of holding 48 dosimeters.

**Measurement conditions** – Irradiations were done with a Varian Truebeam Linear accelerator for both photon and electron measurements. Photon energies of 6 MV, 10 MV, and 15 MV were used as well as electron as at 6 MV and 20 MV. The Truebeam output was determined prior to irradiation using the machine performance check (MPC) output check and the number of monitoring units delivered were adjusted in order to give the desired dose to the dosimeters [7]. For each energy, irradiations were conducted at 100 cm SSD, with the dosimeters at dmax, and 6 cm of solid water for backscatter. Photons used a 10x10 cm field size and electrons used the 10x10 cm cone with a standard 10 cm cutout.

**Batch calibration** – Each dosimeter is an independent device which may have slight differences in performance do to manufacturing deviations, crystal inhomogeneities, and other non-uniform defects. The independent response of each detector can be corrected to the mean of the entire sample using a batch calibration process as described in TG191. An initial 50 cGy irradiation was used to establish per-OSLD calibration coefficients by determining the mean count of the batch, $R_\mu$, divided by the signal of each device, $R_{raw}$, as shown in Equation 1. This batch calibration value was used for each dosimeter for all following readings.

$$BC = \frac{R_\mu}{R_{raw}} \quad \text{Equation 1}$$

**Signal fading** –OSLD signals fade over time due to shallow traps spontaneously freeing over time due to thermal excitation and lattice defects. For this reason, it is recommended by the manufacturer to wait at least 10 minutes post irradiation before reading the device. This allows for the signal response to stabilize with most shallow traps recombining leaving the more stable traps for the dosimetry measurement. For most measurements in this study, a 10-minute (+4 min for readout of all dosimeters) wait time was observed post irradiation. A series of readouts were performed on some measurement sets quantify the signal loss as a function of time. This quantification enables correcting myOSLchip readouts for readout times greater than 10 minutes. We tested readouts out to 3 days.

In addition to time dependent signal loss, the act of stimulating the dosimeters in the readout process results in a reduction in signal for each successive reading. Dosimetric accuracy is improved when this is corrected for in each successive readouts by adding in the signal reduction from each previous measurement. This factor was determined by exposing the dosimeters to a known reference dose of 50 cGy and repeatedly reading out the device and fitting a linear curve to the data. The slope of this line is then used to correct the signal for each successive reading for all dosimeters.

**Dose response** – Dose linearity was evaluated with 6 MV photon irradiations from 50-1050 cGy increasing non-linearly in 50, 100, 150, 200, 250, and 300 cGy exposures. . The measurement sequence was irradiation, 10-minute wait, and device readout, followed by irradiation the next dose. Six irradiations, with doses up to 1050 cGy were performed. The batch of dosimeters were then bleached using the erasure for 2 hours. The whole cycle was repeated 2 times for the same batch of detectors resulting in a dose response curve for 0, 1, and 2 erasure cycles. Energy dependence of the dose response was evaluated by irradiating a batch of 5 dosimeters to 100 cGy at dmax for 10 MV, 15 MV photons, 6 MV electrons, and 20 MV electrons.

**Dosimetry testing** – To determine the system reproducibility with all the correction factors, a test was conducted with a group of 5 detectors irradiated to 200 cGy with 6 MV photons, 1.5 cm of solid water for buildup and 6 cm of buildup backscatter at 100 SSD. A 10-minute wait time was observed and each dosimeter was readout and corrected using all the relevant correction factors following the TG-191 formalism.

## Results & Discussion

**Batch Calibration-** The individual response of each detector was determined for 50 dosimeters. Multiplying the counts observed in each detector by the sensitivity factor, with a range of 0.75 to 1.086, mean of 1, and standard deviation of 0.075, corrects the counts to the mean of the batch. The outliers of the batch calibration were not thrown out, however, selecting dosimeters with a sensitivity factor within 1 standard deviation of the mean could improve the performance of the dosimeter.

**Readout signal loss** – A series of 5 dosimeters readout 16 times, normalized to the first reading and averaged show a linear reduction in counts of 1.6% per reading as seen in Figure 4. This readout signal loss factor was used to correct the raw counts for each subsequent measurement where more than 1 reading was taken before erasure.

**Signal fading -** The signal loss follows an inverse decay and stabilizes at a 4% reduction in signal from 10 minutes to 20 hours (Figure 5). Given that the read out of a batch of dosimeters took ~4 minutes, signal fade across the group of OSLDs resulted in up to 1% uncertainty. The signal can be corrected back to the 10-minute reference time post irradiation if the time of exposure is known using this correction curve. During analysis, it was found that 30 minutes post irradiation was a more stable reference time whereby the relative differences between the two cycles were minimized. While 10 minutes is considered sufficient for shallow trap recombination and light output stabilization, 30 minutes or longer leads to more consistent time dependent light output corrections across the two cycles measured. >3 hours, the signal is stable within a few tenths of a percent.

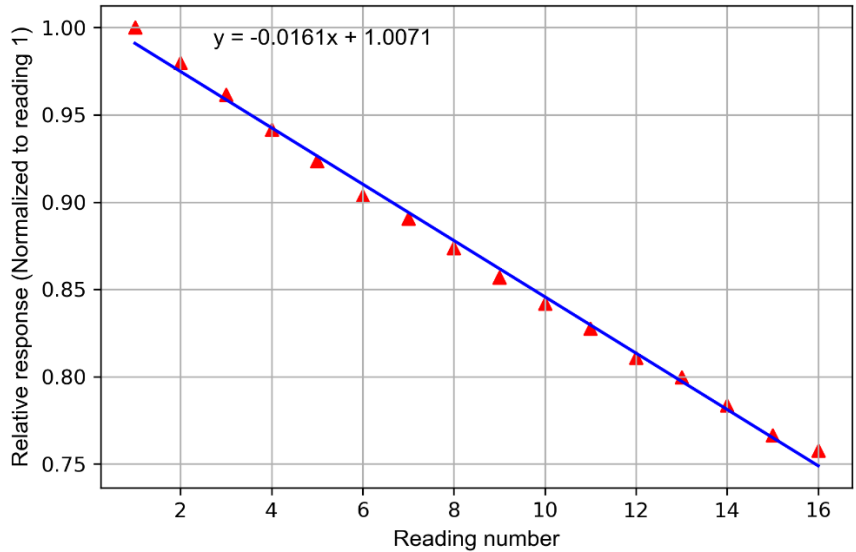

Figure 4. Average signal of 5 dosimeters 10 minutes post exposure. Each dosimeter was readout 16 times consecutively and their signal at each readout number was averaged among the 5 dosimeters. This curve was normalized to the first reading.

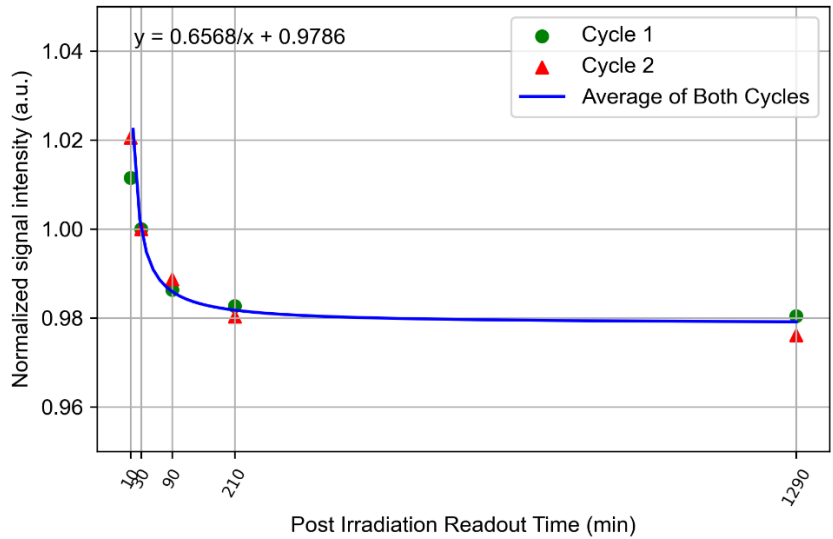

Figure 5. Signal vs. time post 50 cGy irradiation for 5 dosimeters with cycle 1 that had no prior dose, and cycle 2 that was irradiated up to 1050 cGy and erased by spending 2 hours in the eraser. The inverse decay for the two cycles were normalized to 30 minutes which showed a more stable signal output than 10 minutes post exposure. The readout signal loss was corrected for in each reading.

**Dose Linearity -** Each dosimeter was readout 3 times to reduce single reading count variations. The 1.6% signal reduction per-readout was corrected for, as was the per-dosimeter sensitivity factor. The irradiation-readout sequence was repeated for each dose level, yielding a cumulative OSLD dose of 1050 cGy. After the readout of the final dose level, the eraser was used to clear the dosimeters utilizing a high intensity blue LED for two hours to clear the traps in each dosimeter. Each reading sequence was corrected using Equation 2 to normalize the dose response to the first 50 cGy exposure following the TG-191 formalism.

$$k_l(D) = \frac{D_{exp}}{M(D)_{exp}} \Big/ \frac{D_{ref}}{M(D)_{ref}} \quad \text{Equation 2}$$

$D_{ref}$ is the refrence dose of 50 cGy and $M(D)_{ref}$ is the measured signal at the reference dose. $D_{exp}$ and $M(D)_{exp}$ is the dose and measured signal at each exposure level. The irradiation-readout cycle was repeated 2 more times, yielding the dose response shown in Figure 6. The curves in Figure 6 show consistent, linear response up to 750 cGy with variation less than 1%. For one batch of dosimeters, the cycle 2 response for the 1050 cGy dose level was 4% higher than the other batch. As a result, the average cycle 2 variation is larger, but remains less than 2%. Excluding the aberrant cycle 2 reading, the response is linear up to 1050 cGy with less than 1% variation. When looking at the non-normalized, uncorrected counts at each dose level, they are consistently within 4% of other cycles suggesting that erasure and cumulative dose does not significantly impact the light output of the detector. Given the strongly linear dose per cycle, the $N_{D,w}$ calibration factor can be determined at the reference dose of 50 cGy to be 0.00418 cGy/count.

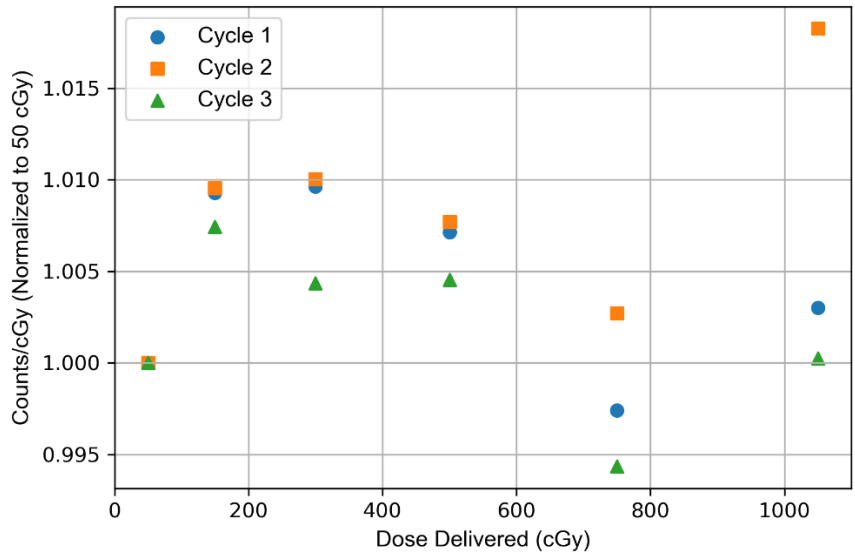

Figure 6. Relative response vs dose delivered for 10 dosimeters averaged, sensitivity corrected, and bleaching corrected for 3 readings per dose level from 50 to 1050 cGy.

**Energy dependance** – The results of the energy dependance of the myOSLchip from 6 to 15 MV photons show a slight energy dependance of 4% for 10 and 15 MV photon beams when compared to 6 MV. However, the difference between 6 MV electrons and 20 MV electrons is negligible when compared to 6 MV photons.

**Dose calculation** - To determine the reliability of the device's dose measurement in a clinical setting, a set of dosimeters were given 200 cGy under the same reference conditions previously stated. Dose is calculated using Equation 3 which is a modified version of the TG-191 fromalism.

$$D_w = M_{corr} \times N_{D,w} \times k_{i_1} \times k_{i_2} \times k_{i_{...}} \quad \text{Equation 3}$$

$D_w$ is the corrected dose, $M_{corr}$ is the background, and individual sensitivity factor corrected counts, and $N_{D,w}$ is the counts to dose conversion factor determined at our reference dose of 50 cGy. The correction factors $k_i$ represent correction factors for signal fading, energy dependency, or off axis factors which were not investigated in this study. The OSLDs reported 199.82±1.34 (0.8%). The readout time for experiment was 10 minutes post irradiation like the reference condition. This shows that with proper characterization of each of these factors, the myOSL system can accurately report the dose delivered in clinical situations.

**Conclusions**

This work has presented the myOSL system as a suitable dosimetry system for clinical use in radiotherapy due to its good dosimetric linearity and accuracy. This device has less than 1% variation up to 750 cGy, increasing to 2% at 1050 cGy with the note of a potential outlier for one of the 1050 cGy batches. The readout as well as time dependent bleaching factors have been well characterized allowing for accurate correction factors for dose conversion. The stable count rate after multiple erasure cycles suggests the device can be erased and reused without significantly altering the dosimetric properties.

**References**

[1] B. J. Nel and S. Perinpanayagam, "A Brief Overview of SiC MOSFET Failure Modes and Design Reliability," *Proc. 5th Int. Conf. -Life Eng. Serv. Cranfield Univ. 1st 2nd Novemb. 2016*, vol. 59, pp. 280–285, Jan. 2017, doi: 10.1016/j.procir.2016.09.025.
[2] P. H. Halvorsen, "Dosimetric evaluation of a new design MOSFET in vivo dosimeter," *Med. Phys.*, vol. 32, no. 1, pp. 110–117, Jan. 2005, doi: 10.1118/1.1827771.
[3] B. Mijnheer, S. Beddar, J. Izewska, and C. Reft, "In vivo dosimetry in external beam radiotherapy," *Med. Phys.*, vol. 40, no. 7, p. 070903, Jul. 2013, doi: 10.1118/1.4811216.
[4] S. F. Kry *et al.*, "AAPM TG 191: Clinical use of luminescent dosimeters: TLDs and OSLDs," *Med. Phys.*, vol. 47, no. 2, pp. e19–e51, Feb. 2020, doi: 10.1002/mp.13839.
[5] E. G. Yukihara and S. W. S. McKeever, "Optically stimulated luminescence (OSL) dosimetry in medicine," *Phys. Med. Biol.*, vol. 53, no. 20, p. R351, Sep. 2008, doi: 10.1088/0031-9155/53/20/R01.
[6] E. G. Yukihara, E. M. Yoshimura, T. D. Lindstrom, S. Ahmad, K. K. Taylor, and G. Mardirossian, "High-precision dosimetry for radiotherapy using the optically stimulated luminescence technique and thin Al2O3:C dosimeters," *Phys. Med. Biol.*, vol. 50, no. 23, p. 5619, Nov. 2005, doi: 10.1088/0031-9155/50/23/014.
[7] M. P. Barnes and P. B. Greer, "Evaluation of the TrueBeam machine performance check (MPC) beam constancy checks for flattened and flattening filter-free (FFF) photon beams," *J. Appl. Clin. Med. Phys.*, vol. 18, no. 1, pp. 139–150, Jan. 2017, doi: 10.1002/acm2.12016.